\documentclass[12pt,psfig]{article}
\newif\iffigs
\figstrue
\newcommand{\be}{\begin{equation}}
\newcommand{\ee}{\end{equation}}
\def\intl{\int\limits}
\def\ms1{{\kern .035em s}}
\def\wiii{${\rm \widetilde{(iii)}}$}
\makeatletter \def\fps@figure{tp} \makeatother
\iffigs
\fi

\def\drawing #1 #2 #3 {
\begin{center}
\setlength{\unitlength}{1mm}
\begin{picture}(#1,#2)(0,0)
\put(0,0){\framebox(#1,#2){#3}}
\end{picture}
\end{center} }
\begin{document}
\setlength{\baselineskip}{.7cm}
\renewcommand{\thefootnote}{\fnsymbol{footnote}}
\sloppy
\begin{center}
\centering{\bf Extreme deviations and applications}
\end{center}

\begin{center}
\centering{U. Frisch $^1$ and D. Sornette $^{2,3}$}
\end{center}
\begin{center}
\centering{
$^1$ Observatoire de la C\^ote d'Azur, CNRS UMR 6529\\
B.P.~4229, F-06304 Nice Cedex 4, France. uriel@obs-nice.fr\\
$^2$ Department of Earth and Space Sciences and Institute of
Geophysics and Planetary Physics,
University of California, Los Angeles, California 90095-1567\\
$^3$ Laboratoire de Physique de
la Mati\`{e}re Condens\'{e}e, CNRS URA 190\\ Universit\'{e} de
Nice-Sophia Antipolis, Parc Valrose, F-06108 Nice, France\\
sornette@naxos.unice.fr
}
\end{center}
\centerline{{\em J. Phys. I France 7, 1155-1171 (1997)}}

\vspace{5mm}
\begin{abstract}
Stretched exponential probability density functions (pdf), having the
form of the exponential of minus a fractional power of the argument,
are commonly found in turbulence and other areas.  They can arise
because of an underlying random multiplicative process.  For this, a
theory of {\em extreme\/} deviations is developed, devoted to the far
tail of the pdf of the sum $X$ of a finite number $n$ of independent
random variables with a common pdf $e^{-f(x)}$. The function $f(x)$ is
chosen (i) such that the pdf is normalized and (ii) with a strong
convexity condition that $f''(x)>0$ and that $x^2f''(x)\to +\infty$ for
$|x|\to\infty$. Additional technical conditions ensure the control of the
variations of  $f''(x)$. The tail behavior of the sum comes then mostly from
individual variables in the sum all close to $X/n$ and the tail of the
pdf is $\sim e^{-nf(X/n)}$. This theory is then applied to products of
independent random variables, such that their logarithms are in the
above class, yielding usually stretched exponential tails.

An application to fragmentation is developed and compared to data from
fault gouges.  The pdf by mass is obtained as a weighted superposition
of stretched exponentials, reflecting the coexistence of different
fragmentation generations.  For sizes near and above the peak size,
the pdf is approximately log-normal, while it is a power law for the
smaller fragments, with an exponent which is a decreasing function of
the peak fragment size.  The anomalous relaxation of glasses can also
be rationalized using our result together with a simple multiplicative
model of local atom configurations. Finally, we indicate the possible
relevance to the distribution of small-scale velocity increments in
turbulent flow.
\end{abstract}

89.90.+n : Other areas of general interest to physicists

\pagebreak
\section{Introduction}
\label{s:intro}

Consider the sum
\be
S_n \equiv \sum_{i=1}^n  x_i,
\label{def}
\ee
where the $x_i$ are independent identically distributed (iid) random
variables with probability density function (pdf) $p(x)$ and mean value
$\langle x\rangle$. The
central limit theorem ensures, with suitable conditions such as the
existence of finite second-order moments or refinements thereof
\cite{GK,Feller}, that as $n\to\infty$ the pdf of
${S_n- n\langle x \rangle \over \sqrt{n}}$ becomes Gaussian. In other words,
the ``typical'' fluctuations of $S_n/n$ around its mean value are Gaussian
and $O(1/\sqrt n)$.  The large deviations theory
is concerned with  events of much lower probability when $S_n/n$ deviates
from its mean value by a quantity
$O(1)$ \cite{Cramer,Varadhan,Ellis,Lanford}. In non-technical terms, for
large $n$,
\be
{\rm Prob}\left[ {S_n \over n} \simeq y \right] \sim e^{n s(y)}  ,
\label{cramera}
\ee
where $s(y)\le 0$ is the Cram\'er function (also called ``rate function'').

In this paper we are concerned with ``extreme deviations'', that is
the r\'egime of {\em finite\/} $n$ and {\em large\/} $S_n=X$. This
r\'egime exists only when the pdf extends to arbitrary large values of
$x$ (i.e.\ has noncompact support). In other words, we are interested
in tail behavior. While it is common in statistics to consider test
probabilities of the order of $1 \%$, much smaller probabilities are
of interest in many areas in which crisis may ensue. If, for instance,
one wishes to investigate whether a chemical substance causes cancer,
one will be interested in very small test probabilities to make a
convincing case. In the field of reliability, failure and rupture, for
instance of industrial plants, very small probabilities are the rule.
Examples are the calculation of the probability of a defect item
passing an inspection system and the calculation of the reliability of
a system. $10^{-6}$ is the probability threshold beyond which the U.S.
Food \& Drug administration considers that any risk from a food
additive is considered too small to be of concern. In the same spirit,
the legal U.S. maximum man-made risk to public is $5\times 10^{-5}$.

The main result about extreme deviations for sums of random variables
is presented in Section~\ref{s:sums}. The relation to large deviations
theory and to other work on extreme deviations is briefly discussed in
Section~\ref{s:large}. Multiplication of random variables is
considered in Section~\ref{s:multiplication}.  Applications are
presented in Section~\ref{s:applications}. The Appendix is devoted to
a rigorous derivation of the main result for sums of independent variables.

\section{Extreme deviations for sums of random variables}
\label{s:sums}

We are interested in the tail behavior for large arguments $x$ of sums of
iid random variables $x_i$, $i=1,2,\ldots.$ We shall only consider large
positive values of $x$. All the results can be adapted {\it mutatis
mutandis\/} to large negative values. We assume that the common probability
distribution of the $x_i$'s has a  pdf, denoted $p(x)$,
which is normalized\,:
\be
\intl_{-\infty}^{\infty}p(x)\,dx=1,
\label{normp}
\ee
and which can be represented as an exponential\,:
\be
p(x) = e^{-f(x)},
\label{pdfexp}
\ee
where $f(x)$ is indefinitely differentiable.\footnote{As we shall see
in the Appendix, this assumption can be relaxed.}
We rule out the case where $f(x)$ becomes infinite at finite $x$;
this would correspond to a distribution with compact support which has
no extreme deviations.

The key assumptions are now listed. All statements involving a limit are
understood to be for $x\to +\infty$.
\begin{itemize}
\item[(i)] $f(x)\to +\infty$ sufficiently fast to ensure the normalization
(\ref{normp}).
\item[(ii)] $f''(x)>0$ (convexity), where $f''$ is the second derivative
of $f$\,.\footnote{This is called log-concavity of the density by Jensen
\cite{Jensen} whose work we shall comment on in Section~\ref{s:large}.}
\item[(iii)] $\lim {f^{(k)}(x)\over
\left(f''(x)\right)^{k/2}} =0$, for $k\ge 3$, where $f^{(k)}$ is the $k$th
derivative of $f$. The $k=3$ instance of (iii) will be denoted by
\wiii.
\end{itemize}
An important consequence of \wiii\,  is
\be
\lim x^2f''(x)=+\infty,
\label{local}
\ee
the proof of which is given in the Appendix (Lemma~1).

We introduce now the pdf $P_n(x)$ of
$S_n = \sum_1^n x_i$ which may be written as a multiple convolution\,:
\be
P_n(x) = \underbrace{\int \cdots\int}_n
e^{-\sum_{i=1}^n f(x_i)}\,\delta\left(x - \sum_{i=1}^n
x_i\right)\,dx_1\cdots dx_n.
\label{solut}
\ee
All integrals are from $-\infty$ to $+\infty$. The delta function expresses
the constraint on the sum.

We shall show that, under assumptions (i)--(iii), the leading-order
expansion of $P_n(x)$ for large $x$ and finite $n\ge 1$ is given by
\be
P_n(x) \simeq e^{-nf(x/n)}
{1\over\sqrt n} \left({2\pi\over f''(x/n)}\right)^{n-1\over 2},\quad {\rm
for} \quad  x
\rightarrow \infty \quad{\rm  and} \quad n \quad {\rm  finite}.
\label{devuf}
\ee
Furthermore, we shall show that the leading contribution comes from
individual terms in the sum which are {\em democratically localized}. By
this we understand that the conditional probability of the $x_i$'s,
given that the sum is $x$, is localized, for large $x$,  near
\be
x_1\simeq x_2\cdots \simeq x_n \simeq {x\over n}.
\label{democratic}
\ee

In this section we give a derivation of this result using a formal
asymptotic expansion closely related to Laplace's method for the asymptotic
evaluation of certain integrals \cite{benderorszag}.\footnote{This method
is sometimes referred to as ``steepest descent'', an inadequate terminology
when $f(x)$ is not analytic.} In the Appendix we shall give a rigorous
proof.

To evaluate (\ref{solut}) for $n\ge 2$, we define new variables
\begin{eqnarray}
h_i &\equiv &x_i -{x\over n}, \quad {\rm for}\,\, i=1,\ldots,n-1
\label{defhi}\\
h_n&\equiv & -\left(h_1+\ldots+h_{n-1}\right),
\label{defhn}
\end{eqnarray}
and the function
\be
g_n\left(x;h_1,\ldots,h_{n-1}\right) \equiv \sum_{i=1}^n f\left({x\over
n}+ h_i\right).
\label{defgn}
\ee
We can then rewrite (\ref{solut}) as
\be
P_n(x)=\underbrace{\int\cdots\int}_{n-1}
e^{-g_n\left(x;h_1,\ldots,h_{n-1}\right)}\, dh_1\cdots dh_{n-1}.
\label{pnhn}
\ee
The function $g_n$ has the following Taylor expansion in powers of the
$h_i$'s\,:
\be
g_n =nf\left({x\over n}\right) +{1\over 2!}f''\left({x\over
n}\right)\sum_{i=1}^n h_i^2 + {1\over 3!}f'''\left({x\over
n}\right)\sum_{i=1}^n h_i^3+\cdots\,.
\label{taylorgn}
\ee
Note the absence of the term linear in the $h_i$'s since, by
(\ref{defhn}), $\sum_{i=1}^n h_i=0$.

If we momentarily ignore the terms of order higher than two in
(\ref{taylorgn}), we obtain for $P_n(x)$ a Gaussian integral the
convergence of which is ensured by the convexity condition (ii). This
integral is evaluated by setting $y=0$ and $\lambda = (1/2)f''(x/n)$
in the identity\footnote{This identity if obtained, after proper
normalization, by evaluating the $n$-fold convolution of a Gaussian
distribution of variance ${1 \over 2 \lambda}$ with itself, which is a
Gaussian of variance ${n \over 2\lambda}$.}
\be
\underbrace{\int\cdots\int}_{n-1}
e^{-\lambda\left[h_1^2+\cdots+h_{n-1}^2 + (y-h_1-\cdots
-h_{n-1})^2\right]}\,dh_1\cdots dh_{n-1} = {1\over\sqrt
n}\left({\pi\over\lambda}\right)^{n-1\over2} e^{-{\lambda\over n}y^2}.
\label{gaussident}
\ee
We thereby obtain the desired asymptotic expression (\ref{devuf}) for
$P_n(x)$.

We now show that higher than second order terms in the taylor
expansion (\ref{taylorgn}) do not contribute to the leading-order
result. The quadratic form $(1/2)f''(x/n)\sum_{i=1}^n h_i^2$ in the
$n-1$ variables $h_1,\ldots,h_{n-1}$ can be diagonalized (it is just
proportional to the square of the Euclidian norm in the subspace
$\sum_{i=1}^n h_i = 0$). One can show by recurrence that it has $n-2$
eigenvalues equal to $(1/2)f''(x/n)$ and one eigenvalue $n$ times
larger. Hence, the Gaussian multiple integral comes from $h_i$'s
which are all $O\left(1/\sqrt{f''(x/n)}\right)$ or smaller. For such
$h_i$'s, it follows from the assumption (iii) that all higher order
terms are negligible for large $x$.  Furthermore, the scatter of the
$x_i$'s around the value $x/n$, measured by the the rms value of the
$h_i$'s is $O\left(1/\sqrt{f''(x/n)}\right)$. By (\ref{local}), this
is small compared to $x$, which proves the {\it democratic
localization\/} property (\ref{democratic}).

We shall also make use of a weaker result obtained by taking the logarithm
of (\ref{devuf}), namely
\be
\ln P_n(x) \simeq - n f(x/ n) ,
\quad {\rm  for} \quad  x
\rightarrow \infty \quad{\rm  and} \quad n \quad {\rm  finite}.
\label{devu}
\ee
This weaker form holds only if
\be
{\ln f''(x/n) \over f(x/n)} \to 0.
\label{weakcond}
\ee

We make a few remarks.
Our derivation is reminiscent of the derivation
of Laplace's  asymptotic formula for integrals of the form
$\int e^{-\lambda f(x)}\,dx$ when $\lambda \to\infty$, as given, e.g.,
in Ref.~\cite{benderorszag}. The main difference is that in Laplace's
method, when  $f$ is Taylor expanded around its minimum, terms
of order higher than two give  contribution smaller by higher and
higher inverse powers of $\lambda$, so that a single small parameter
$1/\lambda$ is enough to justify the expansion, whereas here we made
an infinite number of  assumptions (iii) for all $n\ge3$. Actually,
it will be shown in the Appendix that the sole assumption \wiii\,  with a
slight strengthening of (\ref{local}) is enough to derive the leading-order
term (\ref{devuf}).

It is easily checked that our result is not equivalent to the well-known
fact that the most probable increment ${\Delta x \over \Delta t}$ of a
random walk
conditionned to go from $(x,t)$ to $(x',t')$ is constant and equal to
the average slope ${x'-x \over t'-t}$;  in other words, the most
probable path is then a  straight line, corresponding to a constant reduced
running sum.

The convexity of $f(x)$ at large $x$ is essential for our result
to hold. For instance, pdf's with powerlaw tails $p(x) \propto
x^{-(1+\mu)}$ give $f(x) =(1+\mu) \ln x$ which is concave. The extreme
deviations of the sum $S_n$ are then controlled by realizations where
just one term in the sum dominates. This extends to {\em
arbitrary\/} exponents $\mu$, in this extreme deviations r\'egime, the
well-known result that the breakdown of the central limit theorem for
$\mu < 2$ stems from the dominance of a few large terms in the sum.
The breakdown of democratic localization far in the tail also
happens for pdf's with finite moments of all orders, for example, when
$p(x) \propto x^{-\ln x}$ at large $x$. Here, again the function
$f(x)= \ln^2 x$ is not convex.

The result (\ref{devu}) can  be formally\footnote{Additional assumptions
are then needed to make sure that higher than second-order terms in the
Taylor expansion are not contributing.} generalized to the case  of
dependent variables with nonseparable pdf's $p(x_1,x_2,
...,x_i,...,x_n) =
\exp[-f(x_1,x_2, ...,x_i,...,x_n)]$ where $f(x_1,x_2, ...,x_i,...,x_n)$ is
symmetric and convex. Indeed, ${\partial f \over \partial
x_i}|_{x_1=x_2=..=x_n={S_n \over n}}$ is then independent of $i$ and
the matrix of second derivatives ${\partial^2 f \over \partial x_i^2}$
evaluated at $x_1=x_2=..=x_n={S_n \over n}$ is positive, ensuring that
$f$ is minimum at $x_1=x_2=..=x_n={S_n \over n}$, thereby providing
the major contribution to the convolution integral.

\section{Relation with the theory of large deviations}
\label{s:large}

We now assume, in addition to conditions (i)--(iii) of
Section~\ref{s:sums}, that the characteristic function
\be
Z(\beta) \equiv \langle e^{-\beta x} \rangle =\intl_{-\infty}^\infty
e^{-\beta x}p(x)\,dx
\label{charactt}
\ee
exists for all real $\beta$'s (Cram\'er condition).  Recall that the
Cram\'er function $s(y)$ is determined by the following set of
equations (see, e.g., Refs.~\cite{Varadhan,Lanford,Frisch})\,:
\be
s(y) = \ln Z(\beta) + \beta y  ,
\label{reff}
\ee
\be
{ds(y) \over dy} = \beta .
\label{reffff}
\ee
Hence, $s(y)$ is the Legendre transform of $\ln Z(\beta)$.

Comparison of (\ref{cramera}) with (\ref{devu}) shows that the
Cram\'er function $s(y)$ becomes equal to $-f(y)$ for large $y$. We
can verify this statement by inserting the form $p(x) = e^{-f(x)}$
into (\ref{charactt}) to get $Z(\beta) \sim \int^{\infty}_{-\infty} dx
e^{-\beta x -f(x)}$.  For large $|\beta|$, we can then approximate this
integral by Laplace's method, yielding $Z(\beta) \sim e^{- \min_x
(\beta x +f(x))}$. Taking the logarithm and a Legendre transform, we
recover the identification that $s(y) \to -f(y)$ for large $y$.
Laplace's method is justified by the fact that $|y|
\to \infty$ corresponds, in the Legendre transformation,
to $|\beta| \to \infty$. A number of more precise results are known,
which relate the tail probabilities of random variables to the
large-$y$ behavior of the Cram\'er function. For example, Broniatowski
and Fuchs \cite{Fuchs} give conditions for the asymptotic equivalence
of $s(y)$ (called by them the ``Chernov function'') and of $-\ln
\bar F(y)$ where $\bar F(y)\equiv \int_y^\infty p(x)\,dx$.

A consequence is that the large  and extreme deviations r\'egime  overlap
when taking the two limits $n \to \infty$ and ${S_n \over n} \to
\infty$. Indeed, large deviations theory usually  takes  $n\to \infty$
while keeping ${S_n \over n}$ finite, whereas our extreme deviations
theory takes $n$ finite with $S_n \to \infty$. Our analysis shows
that, in the latter r\'egime, Cram\'er's result already holds for
finite $n$. The true small parameter of the large deviations theory is
thus not ${1 \over n}$ but $\min({1 \over n}, {n \over S_n})$.

A paper by Borokov and Mogulskii \cite{borovkov} contains a result
resembling
somewhat ours. Their eq.~(12) of Section~1 states, in our
notation, that
\begin{equation}
s_n(y) = ns(y/n),
\label{cramersum}
\end{equation}
where $s_n(y)$ is the Cram\'er function for the sum of $n$ independent
and identically distributed copies of a random variable with Cram\'er
function $s(y)$. If we identify the tail of the Cram\'er function with
minus the logarithm of the (tail of the) pdf, their result becomes
identical with (\ref{devu}). However, their result makes no use of
the convexity assumption without which our result will generally not
hold.

Broniatowski and Fuchs \cite{Fuchs} derive a more general but
weaker theorem on the cumulative distribution of $S_n$ for finite $n$,
which ressembles somewhat our result on the {\it democratic
localization\/} property (\ref{democratic}). It is more general
because it is valid for pdf's not obeying the convexity condition
(\ref{local}), for example, the Cauchy distribution. It is weaker
because it states only that there is a number $\alpha_n >0$ such that
\be
\ln {\rm Prob}\, (S_n \geq n x) = \alpha_n [1 + o(1)]
\ln  {\rm Prob}\,(\min(x_1,\ldots,x_n)) \geq x)~,
\label{newpo}
\ee
for $x \to \infty$. Roughly speaking,
(\ref{newpo}) means that the main contributions to the event $S_n \geq n x$
come from the realizations where {\em all\/} variables constituting the sum are
larger than $x$, a  much weaker statement than the property of
{\em democratic localization\/} (\ref{democratic}). 


Jensen \cite{Jensen} also considers the case where $n$ is finite and
the tail probability tends to zero, for particular choices of the pdf.
Jensen is able to show in a few examples that, even though there is no
asymptotics, i.e.\ there is no $n$ tending to infinity, the
saddlepoint expansion allows one to get the correct order of the
probabilities in the tail, using the so-called tilted density
introduced by Esscher \cite{Esscher}. Coupled with the Edgeworth
expansion, this leads to results similar to ours. Our work generalizes
and systematizes these partial results by providing general conditions
of applications, in particular not requiring that $f$ be Taylor
expandable to all orders (see the Appendix).

\section{Multiplications of random variables}
\label{s:multiplication}

Consider the product
\be
X_n = m_1 m_2 .... m_n
\label{productdef}
\ee
of $n$ independent  identically distributed positive\footnote{What
follows is immediately extended to the case of signed $m_i$'s with a
symmetric distribution.} random variables with
pdf $p(x)$. Taking the logarithm of $X_n$, it is clear that we recover the
previous problem (\ref{def}) with the correspondence $x_i \equiv \ln m_i$,
$S_n
\equiv \ln X_n$ and $-f(x) = \ln p(e^x) + x$. Assuming again the set of
conditions (i), (ii) and (iii) on $f$, we can apply
the extreme deviations result (\ref{devu}) which translates into the
following form for the pdf $P_n(X)$ of $X_n$ at large $X$\,:
\be
P_n(X) \sim [p(X^{1 \over n})]^n  ,  \qquad
{\rm  for}\quad X \to \infty\quad {\rm and} \quad n \quad {\rm finite}  .
\label{devuuu}
\ee
[In this section we omit prefactors; this amounts to using
(\ref{devu}) instead of (\ref{devuf}).]
Equation (\ref{devuuu}) has a very intuitive interpretation\,:
the tail of $P_n(X)$ is controlled by the realizations where all terms in
the product are of the same order; therefore $P_n(X)$ is, to leading
order,  just the product
of the $n$ pdf's, each of their arguments being equal to the  common
value $X^{1 \over n}$.

When $p(x)$ is an exponential, a Gaussian or, more generally, of the
form $\propto \exp(-Cx^\gamma)$ with $\gamma>0$, then (\ref{devuuu})
leads to stretched exponentials for large $n$. For example, when $p(x)
\propto
\exp(-Cx^2)$, then $P_n(X)$ has a tail $\propto \exp(-CnX^{2/n})$.

Note that (\ref{devuuu}) can be obtained directly by recurrence.
Starting from $X_{n+1} = X_n x_{n+1}$, we write the
equation for the pdf of $X_{n+1}$ in terms of the pdf's of $x_{n+1}$ and
$X_n$\,:
$$ P_{n+1}(X_{n+1}) = \int_0^{\infty} dX_n P_n(X_n) \int_0^{\infty}
dx_{n+1} p(x_{n+1}) \delta(X_{n+1} - X_n x_{n+1}) = $$
\be
\int_0^{\infty} {dX_n \over X_n} P_n(X_n) p\left({X_{n+1} \over X_n}\right)
 .
\label{geneakfaf}
\ee
The maximum of the integrand occurs for $X_n = (X_{n+1})^{n+1 \over
n}$ at which $X_n^{1 \over n} = {X_{n+1} \over X_n}$.  Assuming that
$P_n(X_n)$ is of the form (\ref{devuuu}), the formal application of
Laplace's method to (\ref{geneakfaf}) then directly gives that
$P_{n+1}(X_{n+1})$ is of the same form.\footnote{Control over
higher-order terms in the asymptotic expansion requires, of course,
the same conditions (i)--(iii) as in Section~\protect\ref{s:sums}.}
Thus, the property (\ref{devuuu}) holds for all $n$ to leading order
in $X$.

Some generalizations are easily obtained. For instance,
for exponential distributions, we can allow for  different characteristic
scales $\alpha_j$ defined by $p_j(x) = \alpha_j
e^{-\alpha_j x_j}$. Eq.~(\ref{devuuu}) then becomes
\be
P_n(X) \sim \exp \biggl( - n \biggl[ X
\prod_{j=1}^n \alpha_j \biggl]^{1 \over n} \biggl) \qquad {\rm for }
\quad X_n >
\prod_{j=1}^n {1 \over \alpha_j}  .
\label{reersulttt}
\ee
Similarly, if $p_j(x) = {2 \over \sqrt{2\pi}
\sigma_j}  e^{-{x_j^2 \over 2 \sigma_j^2}}$, with $x_j \geq 0$, we obtain
\be
P_n(X) \sim \exp \biggl( - {n \over 2} \biggl[ {X^2 \over \prod_{j=1}^n
\sigma_j^2} \biggl] ^{1 \over n} \biggl)
\qquad{\rm for }\quad X_n > \prod_{j=1}^n \sigma_j  .
\label{logsdfnormal}
\ee

\section{Applications}
\label{s:applications}

Considering the simplicity and robustness of the results derived
above, we expect the extreme deviation mechanism to be at work in a
number of physical or other systems. We are thinking in particular of
the application of our result to simple multiplicative processes, that
might constitute zeroth-order descriptions of a large variety of
physical systems, exhibing anomalous pdf and relaxation behaviors.
There is no generally accepted mechanism for their existence and their
origin is still the subject of intense investigation. The extreme
deviations r\'egime may provide a very general and essentially {\it
model-independent\/} mechanism, based on the extreme deviations of
product of random variables.

Fragments are often found to be distributed according to power law
distributions \cite{Redner}\,: In Section~\ref{s:fragment}, we propose
a multiplicative fragmentation model in which the exponent is
controlled by the depth of the cascading process. Anomalous
relaxations in glasses have been largely documented to occur according
to stretched exponentials \cite{glass,Phillips}. In
Section~\ref{s:glass}, we construct a relaxation model based on the
idea that a complex disordered system can be divided into an ensemble
of local configurations, each of them hierarchically ordered.
Stretched exponential pdf are observed in turbulent flow (see, e.g.,
Ref.~\cite{Frisch}) and our extreme deviation theory provides a simple
scenario (Section~\ref{s:turb}). Let us finally mention the question
of stock market prices and their distribution. Here, the very nature
of the pdf's is still debated
\cite{Ghashghaie,challenge}.
While price variations at short time scales (minutes to hours)are
well-fitted by truncated L\'evy laws \cite{Levy}, other alternative
have been proposed \cite{Ghashghaie}. We have found that a stretched
exponential pdf provides an economical and accurate fit to the full
range of currency price variations at the daily intermediate time
scale. We will come back in  future work to document this claim and
to describe the relevance of the multiplicative processes studied
here.

\subsection{Fragmentation}
\label{s:fragment}

Fragmentation occurs in a wide variety of physical phenomena from
geophysics, material sciences to astrophysics and in a wide range of
scales. The simplest (naive) model is to ignore conservation of mass
and to view fragmentation as a multiplicative process in which the
sizes of children are fractions of the size of their parents. If we
assume that the succession of breaking events are independent and
concentrate on a {\em given generation rank\/} $n$, our above result
(\ref{devuuu}) applies to the distribution of fragment size $X$,
provided we take $X$ to zero rather than to infinity. Indeed, the
factors $m_1, m_2, ...,m_n$ are all less or equal to
unity.\footnote{When taking the logarithm, the tail for $X \to 0$
corresponds to the r\'egime where the sum of logarithms goes to $-
\infty$. Although $X \to 0$, is not strictly speaking a ``tail'',
we shall still keep this terminology.} If we take, for example,  $p(m) \propto
\exp\left(-c m^a\right)$ for small $m$, we obtain $P_n(X) \propto
\exp\left(-cn X^{a/n}\right)$. For values of $X$ which are order unity,
large deviations theory applies when $n\to\infty$. This does not, in
general, lead to a log-normal distribution, because central limit
arguments are inapplicable, except in the very neighborhood of the
peak of the pdf of $X$ (see, e.g., Ref.~\cite{Frisch}, Section~8.6.5).

Next, we observe that most of the measured size distribution of
fragments, {\em not conditioned by generation rank}, display actually
power-law behavior $\propto X^{-\tau}$ with exponents $\tau$ between
$1.9$ and $2.6$ clustering around $2.4$
\cite{Turcotte}. Several models have been proposed to rationalize these
observations \cite{Redner,Zhang} but there is no accepted theoretical
description.

Here, we would like to point out a very simple and robust scenario to
rationalize these observations. We again neglect the constraint that
the total mass of the children is equal to that of the parent and use
the simple multiplicative model. Indeed, the constraint of
conservation becomes less and less important for the determination of
the pdf as the generation rank increases. To illustrate what we have
in mind, consider a comminution process in which, with a certain
probablity less than unity, a ``hammer'' repetitively strikes all
fragments simultaneously.  Then the generation rank corresponds to the
number of hammer hits. In real experiments, however, each fragment has
suffered a specific number of effective hits which may vary greatly
from one fragment to the other.  The measurements of the size
distribution should thus correspond to a superposition of pdf's of the
form (\ref{devuuu}) in the tail $X \to 0$.  Recent numerical simulations of
lattice models with disorder \cite{Astrom} show indeed that, for sufficient
disorder, the fragmentation can be seen as a cascade branching process.

Let us now assume that the tail of the size distribution for a fixed
generation rank $n$ is given by (\ref{devuuu}) and that the mean
number $N(n)$ (per unit volume) of fragments of generation rank $n$
grows exponentially\,: $N(n) \propto e^{\lambda n}$ with $\lambda >0$.
It then follows that the tail of the unconditioned size distribution
is given by
\be
 P_{\rm size}(X) \sim \sum_{n=0}^\infty \,[p(X^{1 \over n})]^n e^{\lambda
n}
\sim \int_0^\infty dn \,e^{n \ln p(X^{1 \over n}) +n\lambda}.
\label{psizeapprox}
\ee
Application of Laplace's method in the variable $n$, treated as continuous,
gives a critical (saddle) point
\be
n_\star = -{1\over\alpha}\ln X,
\label{critical}
\ee
where $\alpha$ is the solution of the transcendental equation
\be
\lambda +\ln p\left(e^{-\alpha}\right)  + \alpha e^{-\alpha}
{p'\left(e ^{-\alpha}\right) \over p\left(e ^{-\alpha}\right)} =0.
\label{transcendental}
\ee
The leading-order tail behavior of the size distribution is thus given
by
\be
P_{\rm size}(X) \sim X ^{-\tau}  ,
\label{powerlaw}
\ee
with an exponent
\be
\tau = {1\over\alpha}\left[\ln p\left(e ^{-\alpha}\right)+\lambda\right].
\label{theexponent}
\ee
This solution (\ref{powerlaw}) holds for $\lambda$ smaller than a threshold
$\lambda_c$ dependent on the specific structure of the pdf $p(x)$.
For instance, consider $p(x) \propto \exp\left(-Cx^{\delta}\right)$ for $x \to
0$,
with $\delta > 0$. This corresponds to a pdf going to a constant
as $x \to 0$, with a vanishing slope ($\delta >1$), infinite slope ($\delta
<1$) or finite slope ($\delta = 1$). The equation (\ref{transcendental})
for $\alpha$ becomes ${\lambda \over C} = (1+\alpha \delta) e^{-\alpha
\delta}$. This has a solution only for $\lambda \leq C$, as the function
$(1+x) e^{-x}$ has its maximum equal to $1$ at $x=0$. For $\lambda$
approaching $C$ from below, the exponent of the power law distribution
is given by $\tau = C \delta + O(\sqrt{C-\lambda})$. At the other end
$\lambda \to 0^+$, we get $\tau \to C\delta e$. In between, for $0
\leq \lambda \leq C$, the quantity ${\tau \over C\delta}$ goes
continuously from $e
\approx 2.718$ to $1$. It is interesting that $\tau$ depends on the
parameters
of the pdf $p(x)$ only through the product $C \delta$.

What happens for $\lambda > C$\,? To find out, we return to the
expression (\ref{psizeapprox}) giving the tail of the unconditioned
size distribution and find that the exponential in the integral reads
$e^{n (\lambda - C X^{\delta/ n})}$. In the limit of small fragments
$X \to 0$, the term $X^{\delta/ n}$ is dominated by the large $n$
limit for which it is bounded by $1$. Thus, $\lambda - C X^{\delta/ n}
\leq \lambda - C$. For  $\lambda > C$, the larger $n$ is, the larger
the exponential is, while for $\lambda < C$ there is an optimal
generation number $n_\star$, for a given size $X$, given by
(\ref{critical}). For $\lambda \ge C$, the critical value $n_\star$
moves to infinity.  Physically, this is the signature of a shattering
transition occurring at $\lambda = C$\,: for $\lambda > C$, the number
of fragments increases so fast with the generation number $n$ (as
$e^{\lambda n} > e^{Cn}$) that the distribution of fragment sizes
develops a finite measure at $X=0$.  This result is in accordance with
intuition\,: it is when the number of new fragments generated at each
hammer hit is sufficiently large that a dust phase can appear. This
{\it shattering transition\/} has been obtained first in the context of mean
field linear rate equations
\cite{shatter}.

Consider another class of pdf $p(x) \propto \exp\left(-Cx^{-\delta}\right)$
for $x
\to 0$, with $\delta > 0$. The pdf $p(x)$ goes to zero faster than
any power law as $x \to 0$ (i.e.\ has an essential singularity). The
difference with the previous case is that, as the multiplicative
factor $x \to 0$ occurs with very low probability in the present case,
we do not expect a large number of small fragments to be generated.
This should be reflected in a negative value of the exponent $\tau$.
This intuition is confirmed by an explicit calculation showing that
$\tau$ becomes the opposite of the  value previously
calculated, i.e.\ ${\tau \over C\delta}$ goes continuously from $- e
\approx - 2.718$ to $-1$ as $\lambda$ goes from $0$ to $C$.

In sum, we propose that the observed power-law distributions of
fragment sizes could be the result of the natural mixing occurring in
the number of generations of simple multiplicative processes
exhibiting extreme deviations. This power-law structure is very robust with
respect to the choice of the distribution $p(x)$ of fragmentation ratios,
but the exponent $\tau$ is not universal.
The proposed theory leads us to urge the making  of
experiments in which one can control the generation rank {\it
of each\/} fragment. We then  predict that the fragment
distribution will not be (quasi-)\,universal anymore buton the contrary
characterize better the specific mechanism underlying the fragmentation
process.

The result (\ref{powerlaw}) only holds in the ``tail'' of the
distribution for very small fragments. In the center, the distribution is still
approximately log-normal. We can thus expect a relationship between the
characteristic size or peak fragment size and the tail structure of the
distribution. It is in fact possible to show that the exponent $\tau$ given by
(\ref{theexponent}) is a {\it decreasing\/} function of the peak fragment
size\,:
the smaller is the peak fragment size, the larger will be the exponent (the
detailled quantitative dependence is a specific function of the initial pdf).
This prediction turns out to be verified by the measurements of  particle size
distributions in cataclastic (i.e.\ crushed and sheared rock resulting
in the formation of powder) fault gouge \cite{Sammis}\,: the exponent
$\tau$ of the finer fragments from three different faults (San Andreas, San
Gabriel and Lopez Canyon) in Southern California was observed to be correlated
with the peak fragment size, with finer gouges tending to have a larger
exponent.
Furthermore, the distributions were found to be a power law for the smaller
fragments and log-normal by mass for sizes near and above the peak size.

\subsection{Stretched exponential relaxation}
\label{s:glass}

We would like to suggest a possible application of the stretched
exponential distribution to rationalize stretched exponential
relaxations. {\it A priori}, we are speaking of a different kind of
phenomenon\,: so far we were discussing distributions, while we now
consider the time dependence of a macroscopic variable relaxing to
equilibrium. In contrast to simple liquids where the usual Maxwell
exponential relaxation occurs, ``complex'' fluids \cite{Klinger},
glasses
\cite{glass,Phillips,Palmer}, porous media, semiconductors, etc, have been
found to relax with time $t$ as $e^{-at^{\beta}}$, with $0 < \beta <
1$, a law known under the name Kohlrausch--Williams--Watts law
\cite{glass,Phillips}. Even, the Omori $1/t$ law for aftershock
relaxation after a great earthquake has recently been challenged and
it has been proposed that it be replaced by a stretched exponential
relaxation \cite{Kisslinger}. This ubiquitous phenomenon is still
poorly understood, different competing mechanisms being proposed. An
often visited model is that of relaxation by progressive trapping of
excitations by random sinks \cite{Phillips}. Models of hierarchically
constrained dynamics for glassy relaxations \cite{Palmer} suggest the
relevance of multiplicative processes to account for the relaxation in
these complex, slowly relaxing, strongly interacting materials.  Our
model offers a simple explanation for the difference in $\beta$
measured by the same method on different materials in terms of the
dependence of $\beta$ on the typical number of levels of the hierarchy
as we now show.

We assume that a given system can be viewed as an ensemble of states, each
state relaxing exponentially with a characteristic time scale. Each state
can be viewed locally as corresponding to a given configuration of atoms or
molecules leading to a local energy landscape. As a consequence,
the local relaxation dynamics involves a hierarchy of degrees of freedom up
to a limit determined by the size of the local configuration.
In phase space, the representative point has to
overcome a succession of energy barriers of statistically increasing
heights as time goes on; this is at the origin of the slowing down
of the relaxation dynamics. The characteristic time $t_i$ to overcome
a barrier $\Delta E_i$ is given by the Arhenius factor $t_i \sim
\tau_0 e^{\Delta E_i/kT}$, where $k$ is the Boltzmann constant, $T$
the temperature and $\tau_0$ a molecular time scale. For a succession
of barriers increments, we get that the characteristic time is given
by a multiplicative process, where each step corresponds to climbing
the next level of the hierarchy. In other words, the characteristic
relaxation time of a given cluster configuration is obtained by a
multiplicative process truncated at some upper level. It is important
to notice that our model is fundamentally different from the idea of
diffusion of a representative particle in a random potential with
potential barriers increasing statistically at long times, as in
Sinai's anomalous diffusion \cite{Bouchaud1}. We consider rather that
the system can be divided into an ensemble of local configurations,
each of them hierarchically ordered.

In this simple model, the times $T_n =
\tau_0 (t_1/\tau_0) ... (t_n/ \tau_0)$ are thus log-normally
distributed in their center with stretched exponential tails according to
our
extreme deviation theory. Now, in a macroscopic measurement, one gets
access to
the average over the many different local modes of relaxation, each with a
simple exponential relaxation\,: an observable ${\cal O}$ is thus relaxing
macroscopically as ${\cal O} \sim \langle e^{t \over T_n}
\rangle$, where the average of the observable ${\cal O}$ is carried
out over the distribution of $T_n$.  For large $t$ (compared to the
molecular time scale), Laplace's method gives the leading-order
behavior $$ {\cal O} \sim e^{-at^{\beta}} , $$ with $\beta = {\alpha
\over n + \alpha}$ for a distribution of $T_n$ given by
$e^{-an(T_n/T_0)^{\alpha \over n}}$. In this calculation, we have
assumed that all local configuration clusters are organized
hierarchically according to a fixed number of $n$ levels. We envision
that this organization reflects the local atomic or molecular
arrangement such that the system can be subdivised into a set of
essentially mutually independent local configurations.  These
configurations can tentatively be identified with the locally ordered
structures observed in randomly packed particles \cite{Mosseri},
macromolecules \cite{Macro}, glasses and spinglasses \cite{Mezard}.
The ultrametric structure found to describe the energy landscape of
the spinglass phase of mean field models also leads to a
multiplicative cascade \cite{Bouchaud,Saleur}. Notice that if a system
possesses {\it multiple\/} configuration levels $n$, then by the same
mechanism which in fragmentation led to (\ref{psizeapprox}), the
relaxation becomes a power law instead of a stretched exponential.

The often encountered value $\beta \approx 1/2$ corresponds, in our
model, to the existence of $n \approx \alpha$ levels of the hierarchy.
It is noteworthy that the factor $\alpha$ can be determined
quantitatively from the pdf $p(t_i/\tau_0) \sim \exp[-a
(t_i/\tau_0)^{\alpha}]$ of the multiplicative factors, thus giving the
potential to measure the number of levels of the hierarchy that are
visited by the dynamical relaxation process. This could be checked for
instance in multifragmentation in nuclear collisions, utilizing
techniques sensitive to the emission order of fragments
\cite{nuclear}.

Hierarchical structures are also encountered in evolutionary processes
\cite{evol}, computing architectures \cite{comp} and economic structures
\cite{econ} and, as a consequence, it is an interesting  question whether to
expect dynamical slowing down of the type described above.

\subsection{Turbulence}
\label{s:turb}

In fully developed turbulence, random multiplicative models were
introduced by the Russian school \cite{K62,yaglom,novikovstewart} and
have been studied extensively since. Indeed, their fractal and
multifractal properties provide a possible interpretation for the
phenomenon of intermittency \cite{mandelbrot,parisifrisch} (see also
Ref.~\cite{Frisch}). The pdf's of longitudinal and tranverse velocity
increments clearly reveal a Gaussian-like shape at large separations
and increasingly stretched exponential tail shapes at small separations, as
shown in Figure 1
\cite{meneguzzivincent,Zocchi,grenoble,hollandais,NWLMF96}.

Within the framework of random multiplicative models, our theory
suggests a natural mechanism for the observed stretched exponential tails
at small separations as resulting from extreme deviations in a
multiplicative cascade.  However, this mechanism cannot
account for {\em all\/} properties of velocity increments. For
example, random multiplicative models are not consistent with the
additivity of increments over adjacent intervals. Indeed, the pdf of
velocity increments $\delta v$ cannot become much larger than the
single-point pdf, as it would if the former were $\propto
\exp\left(-C|\delta v|^{\beta}\right)$ with $0<\beta<2$ while the
latter would be close to Gaussian (see the Appendix of
Ref.~\cite{NWLMF96}).  Nevertheless, stretched exponentials could be
working in an intermediate asymptotic range of not too large
increments, the controlling parameter of this intermediate asymptotics
being the separation over which the increment is measured.

\vspace{2mm}

\par\noindent{\bf Acknowledgements}. We have benefitted from
discussions with M.~Blank and with H.~Frisch. This paper is Publication
no.~4711 of the Institute of Geophysics and
Planetary Physics, University of California, Los Angeles.

\newpage
\section*{Appendix} 
\appendix
\renewcommand{\theequation}{\thesection.\arabic{equation}}
\section{Proof of the main results for extreme deviations}
\setcounter{equation}{0}
\label{a:rigorous}

Our aim is to prove (\ref{devuf}) without necessarily assuming that
the function $f(x)$, which defines the pdf of the
individual variables though (\ref{pdfexp}), is Taylor expandable to all
orders. Specifically, we assume that
$f$ is three times continuously differentiable and satisfies the
following conditions when $x\to +\infty$\,:
\begin{itemize}
\item[(i)] $f(x)\to +\infty$ sufficiently fast to ensure the normalization
(\ref{normp}).
\item[(ii)] $f''(x)>0$ (convexity), where $f''$ is the second derivative
of $f$.
\item[\wiii] $\lim {f'''(x)\over
\left(f''(x)\right)^{3/2}}=0$.
\item[(iv)] There exists $C_1>0$ such that, for $x<y$ large enough,
$x^2f''(x)/\left(y^2f''(y)\right) <C_1$.
\item[(v)] There exist $\beta>0$ and $C_2>0$ such that
$x^{2-\beta}f(x)>C_2$ for large enough $x$.
\end{itemize}
Assumptions (i) and (ii) are just the same as made in
Section~\ref{s:sums}.  Assumption \wiii\, is an instance of (iii)
corresponding to the third derivative. Note that nothing is assumed
about higher order derivatives. Assumptions (iv) and (v) are new and will be
seen to be slight strengthenings of a corollary of
\wiii\,. \footnote{There are weaker formulations of (iv) and (v) for which our
results hold, which we do not make explicit here, as they are quite
involved and do not bring any additional insight.}

We begin by proving various lemmas.
\vspace{2mm}
\par\noindent {\bf Lemma 1} {\it
Assumption \wiii~implies}
\be
\lim x^2f''(x)=+\infty.
\label{localrep}
\ee

To prove this result, we start from \wiii\, which may be rewritten as
\be
\lim_{x\to\infty} {d\over dx}{1\over\sqrt{f''(x)}} =0.
\label{d1sqf}
\ee
It follows that for any $\epsilon>0$ there exists $X(\epsilon)$ such
that, for $x>X(\epsilon)$, the argument of the limit in (\ref{d1sqf})
is less than $\epsilon$ in absolute value. We take $X(\epsilon)<y<x$
and apply the mean value theorem to get
\be
\left|\left(f''(x)\right)^{-1/2} -\left(f''(y)\right)^{-1/2}\right|=(x-y)
\left|{d\over d\xi}\left(f''(\xi)\right)^{-1/2}\right|,
\label{mvt}
\ee
with $y<\xi< x$. The rhs of (\ref{mvt}) is less than $(x-y)\epsilon$.
Dividing by $x$ and letting $x\to\infty$, we obtain
\be
\limsup_{x \to\infty} {1\over x \sqrt{f''(x)}} \le \epsilon.
\label{bientot}
\ee
Letting $\epsilon\to 0$, we obtain (\ref{localrep}). QED

\vspace{2mm}
\par\noindent {\bf Lemma 2} {\it
Under assumptions {\rm (ii)},  \wiii~and {\rm (iv)},}
\be
|x-y| = C(f''(x))^{-1/2}
\label{xyclose}
\ee
{\it implies, that}
\begin{equation}
{f''(y)\over f''(x)}\to 1, \quad \hbox{\it for}\,\, x\to +\infty \,\,
\hbox{\it and fixed}\,\, C.
\label{lemme2}
\end{equation}

For the proof, let us first assume that $y<x$.   By the mean
value theorem, we have
\begin{equation}
f''(x)-f''(y) = (x-y)f'''(\xi), \quad {\rm with}\,\, y
<\xi<x.
\label{mvtf2}
\end{equation}
It follows from Lemma~1 and (\ref{xyclose}), that $x/y\to 1$ and thus
$\xi/x\to 1$ as $x\to+\infty$. Dividing (\ref{mvtf2}) by $f''(x)$
and using (\ref{xyclose}), we obtain
\begin{equation}
{f''(x)-f''(y)\over f''(x)} = C{f'''(\xi)\over\left(f''(\xi)\right)^{3/2}}
\left({f''(\xi)\over f''(x)}\right)^{3/2}.
\label{isntit}
\end{equation}
By (iv), the rightmost factor on the rhs is less than $C_1^{3/2}(x/\xi)^3$,
which remains bounded as $x\to+\infty$, while, by \wiii, the leftmost
factor on the rhs tends to zero. Hence, the rhs tends to zero.
This implies (\ref{lemme2}).

For the case $x<y$, (\ref{mvtf2}) holds similarly with $x<\xi<y$. We
then multiply (\ref{mvtf2}) by $(f''(x))^{1/2}/(f''(y))^{3/2}$. The
rhs tends again to zero. If follows that
\be
{f''(x)-f''(y)\over f''(y)}\left({f''(x)\over f''(y)}\right)^{1/2} \to 0,
\label{wellso}
\ee
which implies again (\ref{lemme2}). QED.

\vspace*{2mm}
\par\noindent{\bf Lemma 3} {\it Let $h_i$, $i=1,\ldots, n$ be real
variables, not all vanishing, such that $\sum_{i=1}^n h_i=0$. The
subset of $p\le n-1$ indices $i_j$ such that $h_{i_j}\ge0$ satisfies}
\be
\sum_{j=1}^p h^2_{i_j} \ge {1\over n}\sum_{i=1}^n h^2_i.
\label{hij}
\ee

Let $i_{p+1}$, \ldots, $i_n$ denote the subset of indices such that
$h_{i_j}< 0$. We set $h'_{i_j}=-h_{i_j}>0$, so that
\be
\sum_{j=1}^p h_{i_j}= \sum_{j=p+1}^n h'_{i_j}.
\label{hijhpij}
\ee
We have
\begin{eqnarray}
\sum_{i=1}^n h^2_i &=&\sum_{j=1}^p h^2_{i_j}+\sum_{j=p+1}^n
h'^2_{i_j}\nonumber\\
&\le&\sum_{j=1}^p h^2_{i_j}+ \left(\sum_{j=p+1}^n h'_{i_j}\right)^2
\nonumber\\
&=&\sum_{j=1}^p h^2_{i_j}+ \left(\sum_{j=1}^p h_{i_j}\right)^2
\nonumber\\
&\le& \sum_{j=1}^p h^2_{i_j}+ p \sum_{j=1}^p h^2_{i_j}
\nonumber\\
&\le& n \sum_{j=1}^p h^2_{i_j}.
\label{gosh}
\end{eqnarray}
In deriving (\ref{gosh}), we have used $p\le n-1$ and  the following inequality
for a set  of $p$ nonnegative variables $y_1$, \ldots, $y_p$
\be
(y_1+\cdots +y_p)^2\le p (y_1^2+\cdots+y_p^2).
\label{posineq}
\ee
Lemma~3 follows from (\ref{gosh}). QED.

\vspace*{2mm}
We now turn to the derivation of the main result (\ref{devuf}), rewritten here
in a slightly different form as\,:
\be
\lim_{x\to+\infty} {P_n(x)\over P_n^{\rm as.}(x)} = 1,
\label{psurp}
\ee
where
\be
P_n^{\rm as.}(x)=e^{-nf(x/n)}
{1\over\sqrt n} \left({2\pi\over f''(x/n)}\right)^{n-1\over 2}.
\label{defpas}
\ee

We start from the representation (\ref{pnhn}) of the pdf of the sum of $n$
iid variables as an $(n-1)$-fold integral. The function $g_n$, given by
(\ref{defgn}) can be rewritten as
\be
g_n=nf(x/n) +\sum_{i=1}^n \intl_{x\over n}^{{x\over n}+h_i}dz\intl_{x\over
n}^z f''(y) \, dy.
\label{gnfsec}
\ee
Observe that, by (\ref{defhn}), we have $\sum_{i=1}^n h_i=0$, so that,
ignoring contributions of zero measure, at least one of the $h_i$'s
must be positive.  Furthermore, all the terms involving double
integrals are positive.

The proof goes now as follows. By Lemma~2, the second
derivative $f''(y)$ can be replaced by the second derivative at the
minimizing point $x/n$ as long as all the $h_i$'s are not too large, that
is are in the  set ${\cal A}_H$ defined by
\be
|h_i|\le H=C(f''(x/n))^{-1/2},\quad \hbox{for all}\,\, i.
\label{localized}
\ee
By (\ref{localrep}), (\ref{localized}) expresses that all the
individual random terms in the sum stay within a distance of $x/n$
which is small compared to $x$, that is, what we have called the {\it
democratic localization\/} property.  The substitution of $f''(x/n)$
for $f''(y)$ amounts to using the second-order truncation of the
Taylor series (\ref{taylorgn}) for $g_n$, which leads to $P_n^{\rm
as.}(x)$. It follows from Lemma~2 that the error committed in this
substitution is small for large $x$.

Since $f''(x) > 0$, the contribution of the complementary set
$\overline{{\cal A}_H}$ to the pdf $P_n(x)$, denoted $P_n^{(>H)}(x)$,
is estimated from above by estimating $g_n$ from below, keeping only
the contributions from the subset $i_j$ ($j=1,\ldots, p\le n-1$) of
indices such that $h_{i_j}\ge0$.  We thus obtain
\be
g_n \ge nf(x/n)  +\sum_{j=1}^p\intl_{x\over n}^{{x\over
n}+h_{i_j}}dz\intl_{x\over n}^z f''(y) \, dy.
\label{mingn}\ee
By (iv), for $x/n\le y\le x/n+h_{i_j}$, we have
\be
y^2f''(y)\ge C_1^{-1}(x/n)^2 f''(x/n).
\label{minfsec}
\ee
Using (\ref{minfsec}) in (\ref{mingn}), we obtain
\be
g_n \ge nf(x/n)+C_1^{-1}\left({x\over n}\right)^2 f''(x/n)
\sum_{j=1}^p q(nh_{i_j}/x),
\label{mingnq}
\ee
where
\be
q(\alpha)\equiv \alpha - \ln (1+\alpha).
\label{defq}
\ee
Note that $q(\alpha)= \alpha^2/2 +O(\alpha^3)$ for small $\alpha$ and
$q(\alpha)<\alpha$ for large $\alpha$. Assumption (v) is used to
show that, for large $x$, the overwhelming
contribution to $P_n^{(>H)}(x)$ comes from  $h_{i_j}$'s such that
$nh_{i_j}/x$ is small compared to unity.
Using (\ref{mingnq}) and Lemma~3, we obtain the following estimate
\be
P_n^{(>H)}(x)\le
e^{-nf(x/n)}\underbrace{\int\cdots\int}_{n-1}e^{-{C_1^{-1}\over
2n}f''(x/n)\sum_{i=1}^n h^2_i}\,dh_1\cdots dh_{n-1},
\label{estpnh}
\ee
where the domain of integration is over $\overline{{\cal A}_H}$, so that
at least one of the $|h_i|\ge H=C(f''(x/n))^{-1/2}$. As a consequence, it is
easily checked that the bounding integral is less than $P_n^{\rm as.}(x)$
multiplied  by a factor $O\left(e^{-C^2/n}\right)$, which tends
to zero very quickly for large $C$. This  proves (\ref{psurp}) and the
democratic localization property.

\pagebreak

\noindent FIGURE CAPTIONS

\vskip 1cm

\noindent Figure 1 : pdf of transverse  velocity increments reduced by the rms
velocity at various separations in units of the Kolmogorov dissipation
scale $\eta$. (From Ref.~\protect\cite{NWLMF96}.)

\begin{thebibliography}{99}

\bibitem{GK} B.V. Gnedenko and Kolmogorov, A.N. {\em Limit distributions
for sums of independent random variables}, Addison Wesley, Reading MA
(1954).

\bibitem{Feller} W. Feller, {\it An introduction to probability theory and
its applications}, John
Wiley and sons, New York, vol. II, 1971

\bibitem{Cramer} H. Cram\'er, Sur un nouveau th\'eor\`eme-limite de la
th\'eorie
des probabilit\'es, Actualit\'es Scientifiques et Industrielles 736, 5--23
(1938).

\bibitem{Varadhan} S.R.S. Varadhan, {\it Large Deviations and Applications}
(SIAM,
Philadelphia, 1984).

\bibitem{Ellis} R.S. Ellis, {\it Entropy, Large Deviations and Statistical
Mechanics} (Springer, Berlin, 1985).

\bibitem{Lanford} O.E. Lanford, Entropy and equilibrium states in classical
mechanics, in {\it Statistical Mechanics and Mathematical Problems},
Lect. Notes in Physics 20, 1--113, ed. A. Lenard, Springer, Berlin, 1973.

\bibitem{Jensen} J.L. Jensen, {\it Saddlepoint Approximations}, (Oxford Science
Publications, Clarendon Press, Oxford, 1995).

\bibitem{benderorszag} C. Bender and S.A. Orszag,
{\it Advanced Mathematical Methods for Scientists and Engineers} (McGraw-Hill,
New York 1978).

\bibitem{Frisch} U. Frisch,  {\it Turbulence: the Legacy of
A.N.~Kolmogorov} (Cambridge University Press 1995).

\bibitem{Fuchs} M. Broniatowski and A. Fuchs, Tauberian theorems, Chernoff
inequality and the tail behavior of finite convolutions of distribution
functions, Advances in Mathematics 116, 12--33 (1995).

\bibitem{borovkov}
A.A. Borovkov and A.A. Mogulskii, Large deviations and testing statistical
hypotheses. I. Large deviations of sums of random vectors,
Siberian Advances in Mathematics 2, 52--120 (1992).

\bibitem{Esscher}
F. Esscher, On the probability function in the collective theory of
risk, Skandinavisk Aktuarietidskrift 1932, p.~175.

\bibitem{Redner} S. Redner, Fragmentation, in {\it Statistical models for
the fracture of disordered media}, H.J. Herrmann and S. Roux editors,
Elsevier Science Publishers, (1990); Cheng, Z., and S. Redner, Scaling
theory of
fragmentation, Phys. Rev. Lett. 60, 2450--2453 (1988);
G. Ouillon, D. Sornette, A. Genter and C. Castaing,
The imaginary part of rock jointing, J. Phys. I France 6, 1127-1139 (1996).

\bibitem{glass} G. Gielis and C. Maes, Ergodicity in disordered spin
systems - Stretched exponential relaxation, Europhys. Lett. 31, 1--5
(1995); S.H. Chung and J.R. Stevens, Time-dependent correlation and
the evaluation of the stretched exponential or
Kohlrausch-Williams-Watts function, Am. J. Phys. 59, 1024--1030
(1991); F.  Alvarez, A. Alegria and J. Colmenero, Relationship between
the time-domain Kohlrausch-Williams-Watts and frequency-domain
Havrliak-Negami relaxation function, Phys. Rev. B 44, 7306--7312
(1991).

\bibitem{Phillips} J.C. Phillips, Stretched exponential relaxation in
molecular and electronic glasses, Rep. Prog. Phys. 59, 113
3--1208 (1996).

\bibitem{Ghashghaie} S. Ghashghaie, W. Breymann, J. Peinke, P. Talkner and
Y. Dodge, Turbulent cascades in foreign exchange markets, Nature
381, 767 (1996). S. Ghashghaie, W. Breymann, J.  Peinke and P. Talkner
``Turbulence and financial markets'', in Proceedings European
Turbulence Conference VI, {\sl Advances in Turbulence VI}, 167--170,
S.~Gavrilakis, L.~Machiels and P.A.~Monkewitz, eds. (Kluwer, 1996).

\bibitem{challenge} A. Arn\'eodo, J.-P. Bouchaud, R. Cont,
J.-F. Muzy, M. Potter and D. Sornette, Comment on ``Turbulent cascades in
foreign exchange markets'', cond-mat/9607120) (reply to
Ghashghaie et al. 1996); R. N. Mantegna and H. E. Stanley, ``Stock market
dynamics and turbulence: parallels in quantitative measures of
fluctuation phenomena'', preprint (1995); Turbulence and
financial markets, Nature (Scientific Correspondence), vol. 383, N6601,
587--588 (1996); A. Arn\'eodo, J.-F. Muzy and D. Sornette,
Causal cascade in the stock market from the ``infrared'' to the
``ultraviolet'',
submitted.

\bibitem{Levy}  R. Mantegna and H.E. Stanley,
Scaling behavior in the dynamics of an economic index, Nature 376, 46
N6535, 46--49 (1995).

\bibitem{Turcotte}  D.L. Turcotte, Fractals and fragmentation,
J. Geophys. Res. 91, B2, 1921--1926 (1986).

\bibitem{Zhang} M. Marsili and Y.C. Zhang, Probabilistic fragmentation
and effective power law, Phys. Rev. Lett. 77, 3577--3580. (1996).

\bibitem{Astrom} J. Astrom and J. Timonen, Fragmentation by crack branching,
Phys. Rev. Lett. 78, 3677--3680 (1997).

\bibitem{shatter} D.L. Maslov, Absence of self-averaging in shattering
fragmentation processes, Phys. Rev. Lett. 71, 1268-1271 (1993);
D. Boyer, G. Tarjus and P. Viot, Shattering transition in a multivariate
fragmentation model, Phys. Rev. E 51, 1043--1046 (1995).

\bibitem{Sammis} L.-J. An and C.G. Sammis, Particle size distribution of
cataclastic fault materials from southern California\,: a 3-D study, Pageoph.
143, 203--227 (1994).

\bibitem{Klinger} M.I. Klinger, Glassy disordered systems: topology, atomic
dynamics and localized electron states, Phys. Rep. 165, 275--397 (1988).

\bibitem{Palmer} R.G. Palmer, D.L. Stein, E. Abrahams and P.W. Anderson,
Phys. Rev. Lett. 53, 958 (1984).

\bibitem{Kisslinger} C. Kisslinger, The stretched exponential function as
an alternate model for aftershock decay rates, J. Geophys. Res. 98,
1913--1921 (1993).

\bibitem{Bouchaud1} J.-P. Bouchaud and A. Georges, Anomalous diffusion in
disordered media: Statistical mechanisms, models and physical
applications, Physics Reports 195, 127--293 (1990).

\bibitem{Mosseri} R. Mosseri and J.F. Sadoc, J. Physique Lett. 45, L-827
(1984)

\bibitem{Macro} M. Levitt, Ann. Rev. Biophys. Bioeng. 11, 251 (1982).

\bibitem{Mezard} M. M\'ezard, G. Parisi and M.A. Virasoro, Spinglass Theory
and Beyond, {\em World Scientist Lecture Notes in Physics}, {\em Vol. 9},
1987.

\bibitem{Bouchaud} J.-P. Bouchaud and D.S. Dean,
Aging on Parisi's tree, J. Phys. I France 5, 265--286 (1995).

\bibitem{Saleur} H. Saleur and D. Sornette, Complex exponents and
log-periodic corrections in frustrated systems, J. Phys. I France 6, 327--355
(1996).

\bibitem{nuclear} E.W. Cornell {\it et al.}, Investigating the evolution of
multifragmentating systems with fragment emission order, Phys. Rev. Lett. 77,
4508--4511 (1996).

\bibitem{evol} R. V. O'Neill {\it et al.}, A Hierarchical Concept of
Ecosystems (Princeton University Press, Princeton, N.J. 1986).

\bibitem{comp} B.A. Huberman and M. Kerszberg, J. Phys. A 18, L331 (1985).

\bibitem{econ} Tostesen, E.,  Dynamics of hierarchically clustered
cooperative agents, {\it Cand. Scient. Thesis}, University of
Copenhagen (1995).

\bibitem{K62}
A.N. Kolmogorov,
A refinement of previous hypotheses concerning
the local structure of turbulence in a viscous incompressible fluid
at high Reynolds number,  J.~Fluid~Mech. 13, 82--85 (1962).

\bibitem{yaglom}
A.M. Yaglom, Effect of fluctuations in energy dissipation rate
on the form of turbulence characteristics in the
inertial subrange, Dokl. Akad. Nauk SSSR 166, {49--52} (1966).

\bibitem{novikovstewart}
E.A. Novikov and R.W. Stewart,
{The intermittency of turbulence and the spectrum of energy
dissipation},
Izv., Akad. Nauk SSSR, Ser. Geoffiz., 408--413 (1964).

\bibitem{mandelbrot}
B. Mandelbrot,
{Intermittent turbulence in self-similar
cascades: divergence of high moments and dimension of the carrier},
J.~Fluid~Mech. 62, {331--358} (1974).

\bibitem{parisifrisch}
G. Parisi and U. Frisch,
{On the singularity structure of fully developed turbulence},
in~{\it {Turbulence and Predictability in Geophysical Fluid Dynamics,
Proceed. Intern. School of Physics `E. Fermi', 1983, Varenna, Italy}},
{84--87}, eds.~{M.~Ghil, R.~Benzi \&\,\,G.~Parisi}.
{North--Holland},
{Amsterdam} (1985).

\bibitem{meneguzzivincent}
A. Vincent and M. Meneguzzi,
{The spatial structure and statistical properties
of homogeneous turbulence},
J.~Fluid~Mech. 225, {1--25} (1991).

\bibitem{Zocchi} G. Zocchi, P. Tabeling, J. Maurer and H. Willaime,
Measurement of the scaling of the dissipation at high Reynolds numbers,
Phys. Rev. E 50, 3693--3700 (1994).

\bibitem{grenoble}
H. Kahalerras, Y. Malecot and Y. Gagne, Transverse structure functions
in three-dimensional turbulence.  In {\em Advances in Turbulence\/} VI
(S. Gavrilakis {\it et al.}, eds.), pp.~235--238.  Kluwer (1996).

\bibitem{hollandais}
J.A. Herweijer and W. Van der Water,
Transverse structure functions of turbulence.  In {\em
Advances in Turbulence\/} V (ed. R.~Benzi), pp.~210--216.  Kluwer (1995).

\bibitem{NWLMF96}
A. Noullez, G. Wallace, W. Lempert, R.B. Miles and U. Frisch,
{Transverse velocity increments in turbulent flow
using the RELIEF technique}, J.~Fluid~Mech., in press (1997).



\end{thebibliography}
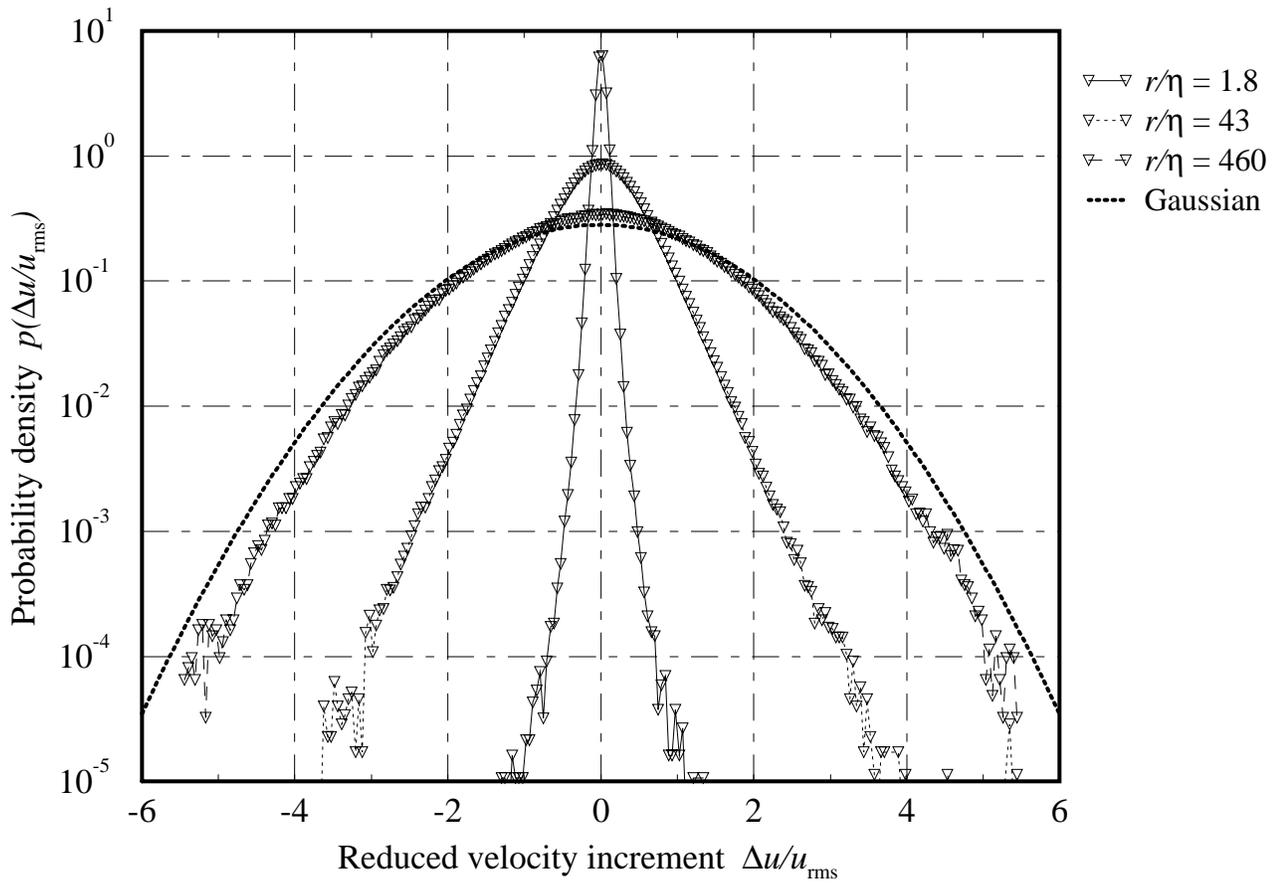
\end{document}